\documentstyle[epsfig]{proceed}

\title[NMA Survey of Active Galaxies]{NMA Survey of CO and HCN Emission from Nearby Active Galaxies}

\author[Kohno, Kawabe, Vila-Vilar\'o]{K. Kohno$^1$\email{kotaro@nro.nao.ac.jp},
R. Kawabe$^1$, \and{} B. Vila-Vilar\'o$^1$}

\institute{$^1$Nobeyama Radio Observatory, Minamimaki, Minamisaku, Nagano,
384-1305, Japan}

\begin{document}
\maketitle

\abstract{
High resolution (a few arcseconds) observations of CO(1$-$0) and HCN(1$-$0) emission
from nearby Seyfert galaxies have been conducted 
with the Nobeyama Millimeter Array.
Based on the observed CO distribution and kinematics, we suggest that
a small scale (a few 100 pc - a few kpc) distortion 
of the underlying potential seems to be necessary for Seyfert activity,
although it is not a sufficient condition.
We also find that Toomre's $Q$ values in the centers of Seyfert galaxies
tend to be larger than unity, suggesting the circumnuclear molecular gas disks
around Seyfert nuclei would be gravitationally stable.
The HCN/CO integrated intensity ratios ($R_{\rm HCN/CO}$)
range over an order of magnitude, from 0.086 to 0.6.
The Seyfert galaxies with high $R_{\rm HCN/CO}$
may have an extended ($r \sim 100$ pc scale) envelope of obscuring material.
The presence of kpc scale jet/outflows might be also related to the
extremely high $R_{\rm HCN/CO}$.
}

\section{Differences between active and non-active galaxies}

Molecular gas has been considered to be a promising fuel source
which may be channeled into the central engines of active galaxies,
and therefore the amount of molecular material in Seyfert galaxies
is expected to be an important clue to understand the nature of
active galactic nuclei.
Surveys of CO emission from Seyfert galaxies with single-dish telescopes
($e.g.$ Maiolino {\it et al.} 1997; Vila-Vilar\'o {\it et al.} 1998)
find, however, no significant differences on the {\it total amount} 
of molecular gas between active and quiescent spirals.
In order to deduce the conditions which are necessary for 
nuclear activity, we have conducted aperture synthesis observations
of CO(1$-$0) and HCN(1$-$0) emission in nearby Seyfert galaxies.
The proximity of the sample ($D < 25$ Mpc; Table 1) 
allows us to perform extensive studies
of their molecular gas distribution, kinematics, and
physical conditions at a few 100 pc scales.

\begin{figure} 
\centering
\includegraphics[width=0.90\textwidth]{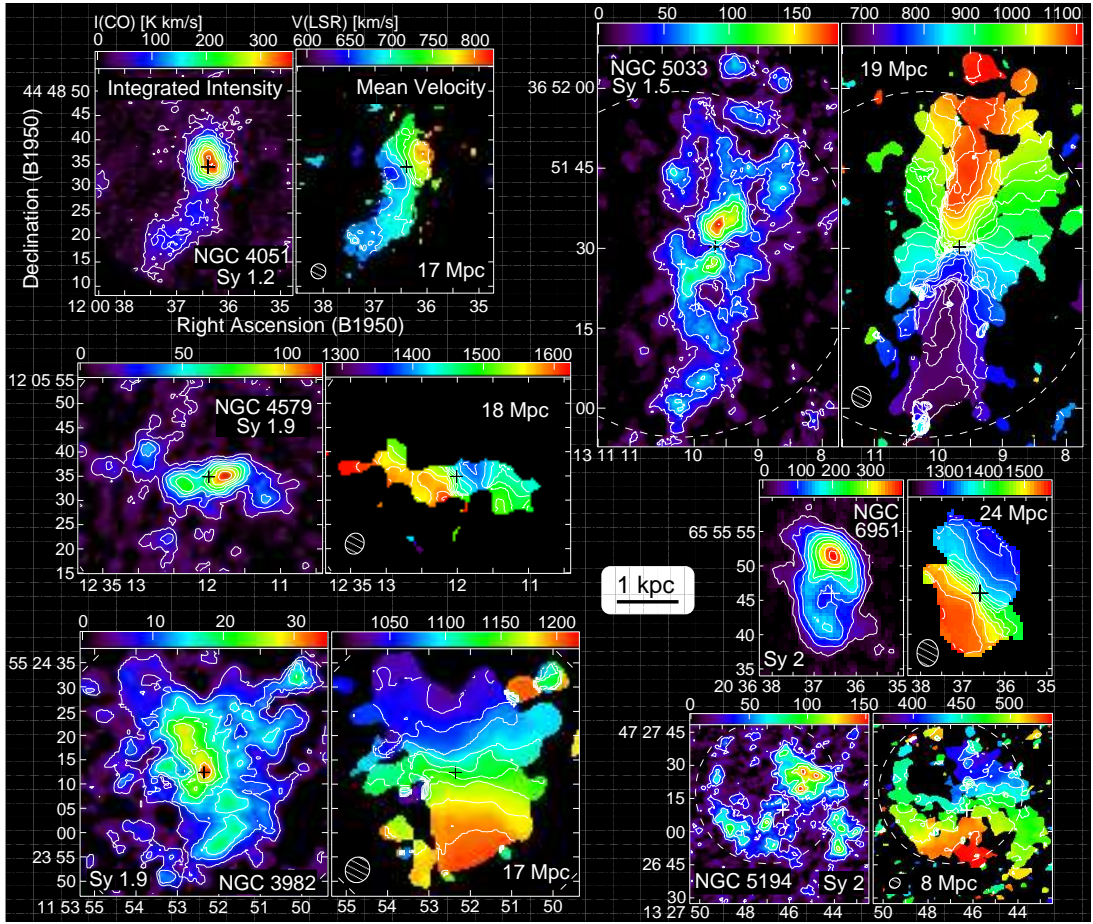}
\caption{Distributions and kinematics of CO emission from the central
regions of nearby Seyfert galaxies.
The size of all CO maps is adjusted to the same linear scale in the figure.
Contour levels are 2 $\sigma$, 4 $\sigma$, 6 $\sigma$, $\cdots$,
for NGC 4051, NGC 4579, NGC 5033, and NGC 5194.
Contour levels for NGC 3982 are 1.5 $\sigma$, 3 $\sigma$,
4.5 $\sigma$, $\cdots$, and 2 $\sigma$, 5 $\sigma$, 8 $\sigma$, $\cdots$, for NGC 6951.
Noise levels ($\sigma$) are 3.3, 19, 6.7, 13, 14, and 15 K km s$^{-1}$
for NGC 3982, NGC 4051, NGC 4579, NGC 5033, NGC 5194, and NGC 6951, respectively.
The cross in each map marks the location of Seyfert nucleus (6 cm radio continuum peak).
}
\label{fig:fig1}
\end{figure}

\section{Distributions and kinematics of molecular gas}

Fig. 1 shows the integrated intensity maps and intensity-weighted mean velocity maps 
of the CO emission of Seyfert galaxies.
The distributions of molecular gas show a wide variety, and
it appears that there are no ``typical'' gas distributions either
in type-1s or type-2s. It should be noted, however, that 
the observed features such as twin peaks (seen in NGC 4579,
NGC 5033, and NGC 6951, for instance)
have been often reported in the central regions of
barred galaxies ($e.g.$ Kenney {\it et al.} 1992),
and all the observed gas distributions and kinematics
seem to be responding to a non-axisymmetric potential
(Kohno {\it et al.} 1999b).
 
It has been claimed that there exist Seyfert galaxies without bars
({\it e.g.} McLeod \& Rieke 1995; Mulchaey \& Regan 1997).
However, very weak bars (a few \% or less in density), 
which may be not easy to detect even in near-infrared bands,
can drastically affect
the behavior of the gas (Wada \& Habe 1992).
Moreover, critical elements of the fuel supply
need not be evident at a large-scale
in the host galaxy (McLeod \& Rieke 1995).
We therefore suggest that small scale (a few 100 pc - a few kpc) distortions
of the underlying potential are necessary for the Seyfert activity,
although it is {\it not} a sufficient condition (Sakamoto {\it et al.} 1998).

\section{Gravitational stabilities of molecular gas}

Self-gravity of molecular gas is another important issue
because circumnuclear molecular gas can lose angular momentum
to fall quickly into the nucleus
if the gas is gravitationally unstable.
The Toomre's $Q$ parameters were therefore computed to examine the
gravitational stabilities of the circumnuclear molecular gas disk 
in Seyfert galaxies (Fig. 2).
We also used CO data in the literature to calculate $Q$ values 
for non-Seyfert galaxies.

Fig. 2 shows that the molecular gas tends to be gravitationally {\it stable}
near the Seyfert nuclei (Kohno {\it et al.} 1999b; Sakamoto {\it et al.} 1998).
This result suggests that instabilities of molecular gas disk
around the nuclei may not cause a further infall of gas in some Seyferts;
if a circumnuclear gas disk becomes gravitationally unstable,
it would result in a burst of star formation there,
as in the case of pure starburst galaxies such as NGC 3504 and NGC 6946.
Note that NGC 4051, which shows $Q < 1$,
has an evidence for recent burst of star formation around the nucleus
(Baum {\it et al.} 1993).

Perhaps our results might be also related to the fact that 
Seyfert nuclei preferentially reside in early type hosts;
it is well known that early type galaxies have smaller gas mass - to -
dynamical mass ratio (Young \& Scoville 1992), and this means
molecular gas in early type galaxies tends to be gravitationally
stable. Even in late type Seyferts such as NGC 5033, 
circumnuclear molecular gas disks may be similar to
those of early type hosts in terms of gravitational stabilities.

\begin{figure} 
\centering
\includegraphics[]{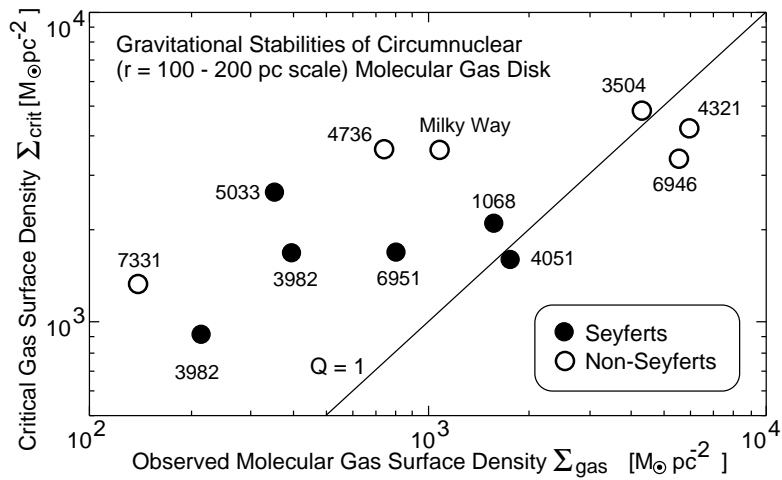}
\caption{Gravitational stabilities of circumnuclear molecular gas disks
measured with the Toomre's $Q$ parameters.
$Q \equiv \Sigma_{\rm crit}/\Sigma_{\rm gas}$,
where $\Sigma_{\rm crit}$ is a threshold gas surface density 
for the onset of instabilities, and $\Sigma_{\rm gas}$ is 
an observed gas surface density. $\Sigma_{\rm crit}$ and 
$\Sigma_{\rm gas}$ are listed in Table 1 as well as $Q$ values.
}
\end{figure}

\section{Physical conditions of molecular gas}

A striking enhancement of HCN(1$-$0) emission has been reported
in the nuclear regions of some Seyfert galaxies
such as NGC 1068 and NGC 5194 (Kohno {\it et al.} 1996 and references therein),
implying extreme physical conditions of the molecular gas near active nuclei
({\it e.g.} Matsushita {\it et al.} 1998). 
Is an extremely enhanced HCN emission common or not in Seyfert galaxies?

The HCN/CO integrated intensity ratios ($R_{\rm HCN/CO}$)
in the observed Seyferts are listed in Table 1 together with $R_{\rm HCN/CO}$
in non-Seyfert galaxies.
The ratios range over an order of magnitude,
from 0.086 to 0.6 among Seyfert galaxies.
$R_{\rm HCN/CO}$ values larger than 0.3 have never been
reported in any non-Seyfert galaxies till date.
We suggest that the observed $R_{\rm HCN/CO}$ in Seyfert galaxies
may be related to the size of {\it dense obscuring material};
NGC 1068 and NGC 5194 have a relatively
extended ($r \sim 100$ pc scale) envelope of obscuring material, 
which consists of very dense interstellar medium and therefore shows 
extremely high $R_{\rm HCN/CO}$.
On the other hand, some Seyferts, NGC 6951 for example, 
may have only a small scale ($r <$ 10 pc) obscuring torus;
in this case, our observing beam
would be too large to find any significant enhancement 
of $R_{\rm HCN/CO}$ associated with the dense obscuring material.

Note that the ``high $R_{\rm HCN/CO}$ Seyferts'',
i.e. NGC 1068, NGC 3079, and NGC 5194, possess kpc scale jet/outflows.
Circumnuclear dense molecular materials may also play a role
in the confinement of large scale jet/outflows (Scoville {\it et al.} 1998).
See Kohno {\it et al.} (1999a, b) against the discussion on the chemical abnormality
of molecular gas in Seyfert galaxies.

 
\begin{table}
\caption{Observed properties of the molecular gas in the central regions of Seyferts and non-Seyferts}
\label{tab:properties}
\begin{center}
\begin{tabular}{llllllc}
\hline \hline
Name & Activity & $\Sigma_{\rm gas}$ & $\Sigma_{\rm crit}$ & $Q$ & $R_{\rm HCN/CO}$ & Ref. \\
\hline
\multicolumn{7}{c}{Seyfert Galaxies (type 1s and 2s)} \\
\hline
NGC 4051 & Sy 1.2 & $1.8 \times 10^3$ & $1.6 \times 10^3$ & 0.90     & $\cdots$ & (1) \\
NGC 5033 & Sy 1.5 & $3.6 \times 10^2$ & $2.7 \times 10^3$ & 7.5      & $0.14$U  & (1) \\
\hline
NGC 1068 & Sy 1.8 & $1.6 \times 10^3$ & $2.1 \times 10^3$ & 1.3      & 0.6     & (2) \\
NGC 3079 & Sy 2   & $2.5 \times 10^3$ & $\cdots$          & $\cdots$ & 0.3     & (3), (1) \\
NGC 3982 & Sy 1.9 & $2.1 \times 10^2$ & $9.1 \times 10^2$ & 4.2      & $0.20$U & (1)  \\
NGC 4579 & Sy 1.9 & $5.2 \times 10^2$ & $\cdots$          & $\cdots$ & $\cdots$ & (1) \\
NGC 5194 & Sy 2   & $4.0 \times 10^2$ & $1.7 \times 10^3$ & 4.2      & 0.56     & (4) \\
NGC 6951 & Sy 2   & $8.2 \times 10^2$ & $1.7 \times 10^3$ & 2.1      & 0.086    & (5) \\
\hline
\multicolumn{7}{c}{Non-Seyfert Galaxies (Starbursts and Post-Starbursts)} \\
\hline
NGC 3504 & Starburst & $4.3 \times 10^3$ & $4.8 \times 10^3$ & 1.1   & 0.3      & (6), (7) \\
NGC 6946 & Starburst & $5.5 \times 10^3$ & $3.4 \times 10^3$ & 0.61  & 0.12     & (1) \\
NGC 4321 & Hot-Spot  & $6.0 \times 10^3$ & $4.2 \times 10^3$ & 0.70  & $\cdots$ & (8) \\
\hline
NGC 4736 & Post-Starburst & $7.6 \times 10^2$ & $3.6 \times 10^3$ & 4.8   & $0.044$U  & (7), (8) \\
NGC 7331 & Post-Starburst & $1.4 \times 10^2$U & $1.3 \times 10^3$ & $9.1$L & $\cdots$ & (9) \\
Milky Way & Young Post-Starburst & $1.1 \times 10^3$ & $3.6 \times 10^3$ & 3.3 & 0.08 & (10), (11) \\
\hline
\end{tabular}
\end{center}
References. --- (1) This work; (2) Helfer \& Blitz 1995; 
(3) Sofue \& Irwin 1992; 
(4) Kohno {\it et al.} 1996; (5) Kohno {\it et al.} 1999a; 
(6) Kenney {\it et al.} 1993; (7) Kohno {\it et al.} 1998; 
(8) Sakamoto {\it et al.} 1998; (9) Tosaki \& Shioya 1997; 
(10) Oka {\it et al.} 1998; (11) Jackson {\it et al.} 1996.

Note. --- (1) Galaxy name; (2) Type of activity;
(3) Observed face-on gas surface density 
$\Sigma_{\rm gas} \equiv 1.36 \Sigma_{\rm H_2}$
in $M_\odot$ pc$^{-2}$, 
adopting $X = 3.0 \times 10^{20}$ cm$^{-2}$ (K km s$^{-1}$)$^{-1}$.
U means upper limit;
(4) Critical gas surface density 
$\Sigma_{\rm crit} \equiv \alpha (\sigma_{v}\kappa/\pi G)$ 
in $M_\odot$ pc$^{-2}$; (5) Toomre's $Q$ value
calculated by $\Sigma_{\rm crit}/\Sigma_{\rm gas}$. L means lower limit; 
(6) HCN to CO integrated intensity radio in brightness temperature scale. 
U means upper limit. See also Kohno {\it et al.} (1999b); 
(7) References.

\end{table}

\bigskip

We are grateful to Dr. T. Helfer for sending her CO data of NGC 1068.
K. K. was financially supported by the Japan Society for the Promotion
of Science.

\end{document}